\documentclass[12pt]{article}

\catcode`\@=11

\global\arraycolsep=2pt
\usepackage{amsbsy,amssymb,latexsym,amsfonts,amsmath}
\usepackage{graphicx,color}

\begin{document}
\begin{flushright}
\parbox{4.2cm}
{CALT 68-2849}
\end{flushright}

\vspace*{0.7cm}

\begin{center}
{ \Large Consistency of local renormalization group in $d=3$}
\vspace*{1.5cm}\\
{Yu Nakayama}
\end{center}
\vspace*{1.0cm}
\begin{center}
{\it California Institute of Technology,  \\ 
452-48, Pasadena, California 91125, USA}
\vspace{3.8cm}
\end{center}

\begin{abstract}
We discuss Weyl anomaly and consistency conditions of local renormalization group in $d=1+2$ dimensional quantum field theories. We give a classification of the consistency conditions and ambiguities in most generality within the power-counting renormalization scheme. They provide many non-trivial constraints on possible forms of beta functions, anomalous dimensions and Weyl anomaly of general $d=1+2$ dimensional quantum field theories.
We perform modest checks of our results in conformal perturbation theories, supersymmetric field theories and holographic computations.
\end{abstract}

\thispagestyle{empty} 

\setcounter{page}{0}

\newpage

\section{Introduction}\label{sec1}
Studies of quantum field theories in curved space-time were originally developed in the context of gravitational physics, such as the probe in black hole geometry and the evolution in cosmology.
However, in recent years, it has been understood that physics of the quantum field theories in curved space-time uncovers far richer structures even if we are ultimately interested in the properties in the flat space-time limit.

In particular, the renormalization group with the space-time dependent cut-off (a.k.a local renormalization group) in the curved space-time and its relation to Weyl anomaly has been playing a significant role in revealing beautiful natures of the landscape of quantum field theories that are connected by the renormalization group flow \cite{Osborn:1991gm}\cite{JO}. It is hard to imagine that the recent progress in our understanding of monotonicity  of the renormalization group flow \cite{Komargodski:2011vj}\cite{Komargodski:2011xv} and the possible equivalence between scale invariance and conformal invariance at the end point of the renormalization group flow \cite{Nakayama:2012nd}\cite{Luty:2012ww}\cite{Fortin:2012hn} were possible without such a formulation that heavily relies on the curved space-time (see e.g. \cite{Nakayama:2013is} for a review).\footnote{To avoid  seemingly pathological counterexamples \cite{Dorigoni:2009ra}\cite{ElShowk:2011gz}, we will assume that our theories can be coupled to gravity with no anomaly in the conservation of the well-defined energy-momentum tensor.}

Moreover, the applicability of the local renormalization group seems to be a foundation of the holographic interpretation of the quantum field theories. While it may be natural to introduce the extra radial direction in holography as the one corresponding to the global renormalization group scale transformation, it is a very particular response of the dual quantum field theories to the local renormalization group that guarantees the full diffeomorphism invariance of the holographic bulk description that treats the field theory directions and the renormalization group direction equally \cite{Lee:2012xba}\cite{Lee:2013dln}. For instance, the invariance under the special conformal transformation rather than the merely scaling transformation plays a crucial role in establishing AdS/CFT correspondence with the full space-time diffeomorphism (rather than foliation preserving diffeomorphism) in the bulk \cite{Nakayama:2012sn}.

The aim of this paper is to understand the implication of the local renormalization group and Weyl anomaly in $1+2$ dimensional space-time. It is typically presumed that the Weyl anomaly only exists in even space-time dimension (see e.g. \cite{Duff:1993wm} for a the historical review of the gravitational contribution to the Weyl anomaly), and it might not be very useful to consider the local renormalization group in odd space-time dimensions. We show this is not the case. By scrutinising the local renormalization group and its consistency conditions in $d=1+2$ dimension, we derive various hidden structures in renormalization group. For instance, we show that beta functions cannot be arbitrary: it must be transverse to various tensors appearing in the Weyl anomaly in $d=1+2$ dimension. We give a classification of the consistency conditions and ambiguities in most generality within the power-counting renormalization scheme. We argue that they provide many non-trivial constraints on possible forms of beta functions and anomalous dimensions of general $d=1+2$ dimensional quantum field theories.

While our main focus is in $d=1+2$ dimension, we hope our systematic approach to the local renormalization group analysis will give comprehensive understanding of this subject in the other dimensions, too. Indeed, we stress that our systematic classification of consistency conditions and ambiguities in local renormalization group will be applicable in any other dimensions with little modifications while the actual expressions may differ in even and odd dimensions. In particular we hope that our discussions on the relatively less known ambiguities in renormalization group will clarify some of the confusions we have encountered in the study of relations between scale invariance and conformal invariance.

The organization of the paper is as follows. In section \ref{sec2}, we begin with the analysis of local renormalization group in the situation where there is no dimensionful coupling constants. Essential features of the local renormalization group in $d=1+2$ dimension will be explained there. Theoretically, we can skip section \ref{sec2} and go directly to section \ref{sec3}, in which we analyse the local renormalization group in most generality within the power-counting renormalization scheme, but we hope that section \ref{sec2} will be pedagogical enough to capture the logic by avoiding too many terms. In section \ref{sec4}, we give some modest checks of our results  in conformal perturbation theories, supersymmetric field theories and holographic computations. In section \ref{sec5}, we conclude with some future perspectives.

We have two appendices. In appendix \ref{appa}, we discuss a possible generalization of the local renormalization group analysis with cosmological constant. In appendix \ref{appb} we collect our conventions and some useful formulae.

\section{Local renormalization group and consistency conditions without mass parameters} \label{sec2}
Let us consider a $(1+2)$ dimensional relativistic quantum field theory originally defined in the flat space-time. In most of the part of this paper, we are implicit about the Wick rotation and work in the Euclidean signature.
The study of the local renormalization group gives non-trivial constraints on possible renormalization group flow. The starting point of the local renormalization group  is to construct the generating functional for correlation functions (i.e. Schwinger vacuum energy functional \cite{Schwinger:1951xk}) by promoting coupling constants $g^I$ to space-time dependent background fields $g^{I}(x)$.
\begin{align}
e^{W[g^I(x)]} = \int \mathcal{D} X e^{-S_0[X] - \int dx^3 g^I(x) O_I(x) + \mathcal{O}(g^2) } \ ,
\end{align}
where $O_I(x)$ are all the (primary) operators in the theory (we will also discuss various tensorial operators below).\footnote{There is a small caveat here. If $O_I(x)$ (rather than its space-time integral) is not well-defined, the promotion of the coupling constants to background fields may not be possible. A famous example is the Chern-Simons interaction. At the same-time, in such situations, there is a topological obstruction so that the renormalization of such coupling constants are very much constrained (e.g. only 1-loop shift in Chern-Simons theory). We can simply treat such coupling constants as external fixed parameters in the following argument. In particular there is no associated Weyl anomaly.}

The $\mathcal{O}(g^2)$ higher order terms in the definition of the renormalized Schwinger functional contain some arbitrariness related to contact terms and scheme dependence, which we will dwell on later. However, at this point, we should mention that there are two types of important background fields whose structure of the contact terms may be constrained by requiring the relevant Ward-Takahashi identities.
The first one is the background metric $\gamma_{\mu\nu}(x) = \eta_{\mu\nu} + h_{\mu\nu}(x) + \cdots$ (here $\eta_{\mu\nu}$ is the flat space-time metric) that naturally couples with the energy-momentum tensor as $h_{\mu\nu} T^{\mu\nu} + O(h^2)$. The arbitrariness for the coupling to the background metric is reduced by requiring that the vacuum energy functional $W[\gamma_{\mu\nu}(x), g^I(x)]$ is diffeomorphism invariant with respect to the background metric $ds^2 = \gamma_{\mu\nu}(x) dx^{\mu} dx^{\nu}$. Still, it does not fix the arbitrariness entirely because there are higher curvature corrections such as the $\xi R \phi^2$ term in scalar field theories with $R$ being the Ricci scalar which cannot be fixed without further assumptions (e.g. Weyl invariance or supersymmetry). We could also add the local counterterms constructed out of metric which is diffeomorphism invariant.

The second important example is the background vector fields $a_{\mu}(x)$ that couple to not-necessarily-conserved vector operators $J^\mu(x)$.  Generically, the vector operators $J^\mu$ are not conserved due to the source terms $g^I(x)O_I(x)$ in the interaction. In order to systematically implement the broken Ward-Takahashi identities for the vector operators $J^\mu$, it is convenient to introduce the compensated gauge transformations for the source of the violation such as $g^I(x)$ so that the vacuum energy functional $W[\gamma_{\mu\nu}(x), g^I(x), a_{\mu}(x)]$ is invariant under the compensated gauge transformation:
\begin{align}
\delta a_\mu(x) &= D_{\mu} w(x) \cr
\delta g^I(x) &= -(wg)^I(x) \ . \label{compensg}
\end{align}
Here we assume that the ``free part" of the action $S_0[X]$ has the symmetry $\mathcal{G}$ and the background gauge fields $a_{\mu}(x)$ lies in the corresponding Lie algebra $\mathfrak{g}$. The coupling constants $g^I(x)$ form a certain representation under $\mathcal{G}$. 
We will denote the covariant derivative $D_{\mu} = \partial_\mu + a_{\mu}$ and the field strength $f_{\mu\nu} = \partial_\mu a_\nu - \partial_\nu a_{\mu} + [a_{\mu},a_{\nu}]$ as usual in the matrix notation. When the covariant derivative acts on tensors, they must contain the additional space-time connection.
This compensated gauge invariance plays a significant role in understanding the  importance of operator identities in the local renormalization group analysis \cite{Osborn:1991gm}\cite{JO}.

The crucial assumption in the following is that the Schwinger vacuum energy functional is finitely renormalized (renormalizability assumption). Theoretically this assumption is a great advantage because varying the renormalized Schwinger functional automatically takes into account the renormalization of the composite operators.\footnote{This is a chicken or egg problem in the actual computation because we have to renormalize the infinite set of operators with derivatives to construct the renormalized vacuum energy functional after all. However, the general structure of the renormalization group flow is more transparently seen in just declaring its existence.} The renormalization group equation for this Schwinger functional, whose study is the main goal of this paper, is known as the local renormalization group equation  \cite{Osborn:1991gm} because we perform the space-time dependent change of coupling constants as well as renormalization scale. This has a huge advantage in discussing the conformal invariance (rather than merely scale invariance) because it directly provides the response to the non-constant Weyl transformation. 

Throughout this section, we concentrate on the so-called massless renormalization group flow in which we have no dimensionful coupling constants.
Without any dimensionful coupling constant at hand, the local renormalization group operator can be expressed as
\begin{align}
\Delta_{\sigma} &= \int d^3x \sqrt{|\gamma|} \left( 2\sigma \gamma_{\mu\nu} \frac{\delta}{\delta \gamma_{\mu\nu}} + \sigma \beta^I \frac{\delta}{\delta g^I}  \right. + \left. \left( \sigma \rho_I D_{\mu} g^I - (\partial_\mu \sigma) v \right)\cdot \frac{\delta}{\delta a_{\mu}} \right) \ . 
\end{align}
In the subsequent sections, we will study further generalizations with dimensionful coupling constants. The assumption of the renormalizability is equivalent to the claim that the Schwinger functional is annihilated by $\Delta_{\sigma}$ up to the Weyl anomaly that is a local functional of the renormalized sources. 

The each term in $\Delta_{\sigma}$ has a simple interpretation. The first term $2\sigma \gamma_{\mu\nu} \frac{\delta}{\delta \gamma_{\mu\nu}}$  generates nothing but the Weyl rescaling of the metric by the Weyl factor $\sigma(x)$: $\delta_{\sigma}\gamma_{\mu\nu}(x) = 2\sigma(x) \gamma_{\mu\nu}(x)$. The renormalization of the coupling constants introduce additional running of the coupling constants under the change of the local scale transformation:
$\beta^I$ is the scalar beta function for the corresponding operator $O_I$ which is necessary to cancel the divergence appearing in the coupling constant renormalization for $g^I$.   Less familiar terms $\rho_I$ and $v$ are related to the renormalization group running for the vector background source $a_{\mu}$. We emphasize that once the coupling constant $g^I(x)$ is space-time dependent, we have extra divergence in relation to vector operators that must be cancelled by renormalizing the background vector fields $a_{\mu}$. Even in the flat space-time limit, such effects are actually visible as the renormalization of the composite vector operators.

The invariance of the Schwinger functional under the local renormalization group (up to anomaly) corresponds to the trace identity
\begin{align}
T^{\mu}_{\ \mu} = \beta^I O_I + (\rho_I D_\mu g^I) \cdot J^\mu + D_\mu (v\cdot J^\mu) + A_\mathrm{anomaly} \ \label{traceiden}
\end{align}
from the definition of the renormalized composite operators:
\begin{align}
2\frac{\delta}{\delta \gamma^{\mu\nu}(x)} W &= -\langle T_{\mu\nu}(x) \rangle  \cr
\frac{\delta}{\delta g^I(x)} W &= - \langle O_I(x) \rangle \cr 
\frac{\delta}{\delta a_{\mu}(x)}W &= - \langle J^\mu(x) \rangle \ .
\end{align}
These relations are typically known as the Schwinger (quantum) action principle \cite{Schwinger:1951xk}. In our local renormalization group approach, it simply gives the definition of the renormalized composite operators.  
The last term $A_\mathrm{anomaly}$ in \eqref{traceiden} is a $c$-number that depends on the space-time dependent coupling constants or background fields, usually known as Weyl anomaly (or trace anomaly) that we will discuss below.

As we will discuss in more detail in section \ref{sec2.2}, the Schwinger functional must be invariant under the compensated gauge transformation \eqref{compensg}: 
\begin{align}
\Delta_{w} W[\gamma_{\mu\nu},g^I, a_{\mu}]  = \int d^3x \sqrt{|\gamma|} \left(D_\mu w \cdot \frac{\delta}{\delta a_{\mu}} - (wg)^I \frac{\delta}{\delta g^I} \right) W[\gamma_{\mu\nu},g^I, a_{\mu}] = 0 \ \label{gaugetrans}
\end{align}
for any Lie algebra element $w \in \mathfrak{g}$ that generates the compensated symmetry $\mathcal{G}$, 
so the local renormalization group operator can be equivalently rewritten as
\begin{align}
\Delta_{\sigma} &= \int d^3x \sqrt{|\gamma|} \left( 2\sigma \gamma_{\mu\nu} \frac{\delta}{\delta \gamma_{\mu\nu}} + \sigma B^I \frac{\delta}{\delta g^I}  \right. + \left. \left( \sigma \hat{\rho}_I D_{\mu} g^I \right)\cdot \frac{\delta}{\delta a_{\mu}} \right) \ , \label{lrgrv} 
\end{align}
when we act on the gauge invariant $W[\gamma_{\mu\nu},g^I, a_{\mu}]$, where
\begin{align}
B^I &= \beta^I - (vg)^I \cr
\hat{\rho}_I &= \rho_I + \partial_I v \ .
\end{align}
In the language of the trace identity, rewriting here corresponds to the use of the operator identity or the equation of motion\footnote{This equation may seem to assume implicitly that the tree level equations motion are the same as the renormalized ones. Depending on the renormalization scheme, it may not be the case and it is possible to have corrections such that $(wg)^I$ is effectively replaced by $(Xwg)^I$, where $X = 1+ O(g^I)$ now contains the higher order corrections. Similar ambiguities appear in section \ref{sec2.2} (Class 2 ambiguity). Such a possibility is unavoidable in $d=1+3$ dimension due to possible gauge anomaly in the right hand side of \eqref{gaugetrans}. We do not expect the gauge anomaly in $d=1+3$ dimension, but we may have (fractional) Chern-Simons counterterms we will discuss later. In any case, after rewriting it as in \eqref{lrgrv} with whatever renormalized operator identity we have in the theory, there will be no significant difference in the following.}
\begin{align}
v \cdot D_\mu J^\mu = -(vg)^I O_I \ 
\end{align}
so that we have the equivalent expression \cite{Osborn:1991gm}
\begin{align}
T^{\mu}_{\ \mu} = B^I O_I + (\hat{\rho}_I D_\mu g^I) \cdot J^\mu + A_\mathrm{anomaly} \ . \label{trirv}
\end{align}

Although the physics does not change with the gauge (for the background fields) which we choose, we will mostly stick to the conventional choice \eqref{lrgrv} and \eqref{trirv} in the following. This choice has a great advantage in the flat space-time limit because $B^I = 0$ directly implies the conformal invariance (i.e. $T^{\mu}_{\ \mu}|_{\gamma_{\mu\nu} = \eta_{\mu\nu}} = 0$). If we used the other choice, we would have to keep track of both $\beta^I$ and $v$ to compute $B^I = \beta^I - (vg)^I$ in order to discuss the conformal invariance. For this reason, it is most convenient \cite{JO}\cite{Nakayama:2012nd} to define the renormalization group equation for the running background source fields by
\begin{align}
\frac{dg^I}{d\sigma} &= B^I \cr
\frac{da_{\mu}}{d\sigma} &= \hat{\rho}_I D_{\mu} g^I \ . 
\end{align}
Again, we could evolve the coupling constants in whatever gauge we like (i.e. $\frac{dg^I}{d\sigma} = \beta^I$), and the physics does not change. However, the conformal invariance at the fixed point would be disguised.

In the flat space-time limit, the physical meaning of these equations can be summarized as the (massless) Callan-Symanzik equation or Gell-Mann Low equation:
\begin{align}
\left(\frac{\partial}{\partial \log\mu} + \beta^I \frac{\partial}{\partial g^I} \right) W[\gamma_{\mu\nu}(x) = \eta_{\mu\nu}, g^I(x) = g^I, a_\mu(x) = 0] = 0 \ ,
\end{align}
where $\mu$ is the space-time independent renormalization scale.
Here the generator of the constant scaling transformation by the metric is replaced by the change of the renormalization scale $\mu$ from the dimensional counting.
Note that (A) the contribution from the source term anomaly $A_{\mathrm{anomaly}}$ is gone, and (B) the total divergence terms $D^\mu J_\mu$ in the trace identity do not contribute (at least except for possible contact terms) so that one can replace $B^I$ with $\beta^I$, which makes it harder to keep track of these terms in the flat space-time renormalization \cite{Nakayama:2012nd}.\footnote{Note that due to the contact terms, we do have to keep track of the wave-function renormalization factor and equation of motion terms if we compute the higher point (integrated) correlation functions. These contact terms will be different when we use $\beta^I$ functions than when we use $B^I$ functions.}

In $d=1+2$, without any mass parameter, the allowed structure of the Weyl anomaly is limited from the power-counting renormalization scheme that we assume. Up to total derivatives, we have the anomalous Weyl variation
\begin{align}
A_{\sigma} & = \Delta_\sigma W|_{\mathrm{anomaly}}  \cr
&= \int d^3 x \sqrt{|\gamma|} \sigma \left( \epsilon^{\mu\nu\rho}  C_{IJK} D_{\mu} g^I D_{\nu} g^J D_{\rho} g^K + \epsilon^{\mu\nu\rho} f_{\mu\nu} \cdot C_I \cdot D_{\rho} g^I \right) \ . \label{3danomaly}
\end{align}
Here $C_{IJK}$ maps $(R_I \otimes R_J \otimes R_K) \to \mathbf{R}$,\footnote{We always choose $C_{IJK}$ to be antisymmetric with respect to permutations of $IJK$: $C_{IJK} = C_{[IJK]}$. See appendix \ref{appb} for our convention of antisymmetric symbol.} and $C_I$ maps $(\mathrm{adj} \otimes R_I) \to \mathbf{R}$ under the compensated symmetry group $\mathcal{G}$.
Equivalently, we have the trace anomaly from the space-time dependent coupling constants:
\begin{align}
T^{\mu}_{\ \mu} |_{\mathrm{anomaly}} = A_{\mathrm{anomaly}} =   -\epsilon^{\mu\nu\rho} C_{IJK} D_{\mu} g^I D_{\nu} g^J D_{\rho} g^K - \epsilon^{\mu\nu\rho} f_{\mu\nu} \cdot C_I \cdot D_{\rho} g^I \ . \label{traceanomalymassless}
\end{align} 
Note that CP must be broken due to the appearance of the Levi-Civita tensor $\epsilon^{\mu\nu\rho}$ to obtain this non-trivial trace anomaly. We also notice that for a constant scale transformation (i.e. $\partial_\mu \sigma = 0$), we have the equivalence relations $C_{IJK} \sim C_{IJK} + \partial_{[I} \Lambda_{JK]}$ and $C_{I} \sim C_{I} + \partial_I \Lambda$ thanks to possible integration by part. Thus, the constant scale anomaly is weaker than the Weyl anomaly in such a situation (see e.g. \cite{Nakayama:2012sn} for a similar argument in relation to holography).

\subsection{Consistency condition} \label{sec2.1}
So far, we have introduced various beta functions and anomalous Weyl variations  for space-time dependent sources. The important observation is that there exist non-trivial consistency conditions they must satisfy. In this subsection, we discuss such consistency conditions in a systematic way. 

We first propose that there are two distinct classes of consistency conditions from the integrability of the local renormalization group.
\begin{itemize}
\item
Class 1 consistency condition: Integrability conditions for the local renormalization group transformation operator
\item
Class 2 consistency condition: Integrability conditions for the Weyl anomaly
\end{itemize}
Both of them are based on the requirement that the local renormalization group (or Weyl transformation) is Abelian:
\begin{align}
[\Delta_\sigma, \Delta_{\tilde{\sigma}}] = 0 \ .
\end{align}
This is known as the Wess-Zumino consistency condition \cite{Osborn:1991gm}.

Class 1 consistency condition comes from the general property of the local renormalization group operator $\Delta_{\sigma}$, and it does not depend on the specific form of the Weyl anomaly. Therefore, Class 1 consistency condition is more or less independent of the space-time dimension $d$ while we focus on the $d=1+2$ in this paper. The requirement of the commutation relation
\begin{align}
[\Delta_\sigma, \Delta_{\tilde{\sigma}}] W[\gamma_{\mu\nu}, g^I , a_{\mu}] = 0
\end{align}
on any (local or non-local) functional $W[\gamma_{\mu\nu}, g^I, a_{\mu}]$, we must demand\footnote{More precisely, the integrability condition must be only true for the functional $W[\gamma_{\mu\nu}, g^I, a_{\mu}]$ that is consistent with the local renormalization group so at this stage it may not be necessarily true for arbitrary functionals. As we will discuss, however, we can always add local counterterms on $W[\gamma_{\mu\nu}, g^I, a_{\mu}]$, so the following requirement that can be obtained from the action on the local functional is certainly necessary for our purpose.}
\begin{align}
 -\int d^3x\sqrt{|\gamma|} (\sigma \partial_\mu \tilde{\sigma} - \tilde{\sigma} \partial_\mu \sigma) B^I \hat{\rho}_I \cdot \frac{\delta}{\delta a_{\mu}} = 0 \ ,
\end{align}
or
\begin{align}
B^I \hat{\rho}_I = 0 \ , \label{rhocons}
\end{align}
which shows a transversal condition of the beta functions. Note that this condition is same as the one we found in $d=1+3$ dimension \cite{Osborn:1991gm}, which played an important role in deriving perturbative strong $a$-theorem with non-trivial vector operators.

On the other hand, Class 2 consistency condition comes from the anomalous terms $A_{\mathrm{anomaly}}$ (or its integrated form $A_{\sigma}$) in the local renormalization group transformation, and therefore the following conditions are unique to $d=1+2$ dimension. 
The Wess-Zumino consistency condition on the anomalous variation demands
\begin{align}
\Delta_{\tilde{\sigma}} A_{\sigma} = \Delta_{\sigma} A_{\tilde{\sigma}} \ .
\end{align}
by recalling the definition of the anomaly $A_{\sigma} = \Delta_\sigma W$.
By substituting the available form of the anomaly \eqref{3danomaly}, and using the variational formula
\begin{align}
\Delta_{\sigma} D_\mu g^I &= \partial_\mu \sigma B^I + \sigma(\partial_J B^I + (\hat{\rho}_J g)^I) D_\mu g^J \cr
\Delta_{\sigma} f_{\mu\nu} &= \sigma( (f_{\mu\nu} g)^I \hat{\rho}_I + (\partial_I \hat{\rho}_J - \partial_J \hat{\rho}_I) D_{\mu} g^I D_{\nu} g^J ) \cr
& + \partial_\mu \sigma \hat{\rho}_I D_\nu g^I - \partial_\nu \sigma \hat{\rho}_I D_\mu g^I \ ,
\end{align}
we obtain the consistency condition from terms proportional to $D_{\mu}g^I D_\nu g^J$ and $f_{\mu\nu}$ as
\begin{align}
3B^I C_{IJK} + \hat{\rho}_J C_K - \hat{\rho}_{K} C_J & = 0 \cr
B^I C_I & = 0 \ . \label{consistenccc}
\end{align}
Note that contracting the first equation with $B^J$ requires the second equation from Class 1 consistency condition $B^I \hat{\rho}_I = 0$. Again, the consistency conditions require that the beta functions must satisfy certain transversality conditions.
With the same logic, Osborn \cite{Osborn:1991gm} derived Class 2 consistency conditions for the Weyl anomaly in $d=1+1$ and $d=1+3$ dimension, among which he obtained
\begin{align}
B^I \partial_I \tilde{A} = - g_{IJ} B^I B^J \label{atheorem}
\end{align}
with a certain ``metric" $g_{IJ}$ and a potential function $\tilde{A}$ on the coupling constant space. This equation provided a foundation of the perturbative proof \cite{Osborn:1991gm} of $c$-theorem \cite{Zamolodchikov:1986gt} in $d=1+1$ and $a$-theorem \cite{Cardy:1988cwa}\cite{Komargodski:2011vj} in $d=1+3$, where $\tilde{A}$ is identified as the interpolating $a$-function along the renormalization group flow. Our results do not directly give the analogous monotonicity results in $d=1+2$ dimension, but they still show non-trivial constraints on the renormalization group.

\subsection{Ambiguity} \label{sec2.2}
The renormalization group has intrinsic ambiguities typically known as scheme dependence. The use of the local renormalization group leads to a classification of such ambiguities in a systematic manner. A well-known scheme dependence (e.g. various subtraction scheme in dimensional regularization) is understood as a particular subclass (Class 2) of the ambiguities we will discuss in this subsection. Broader classes of ambiguities play a significant role in understanding composite operator renormalization and the operator mixing such as energy-momentum tensor.

We have three distinct classes of ambiguities in local renormalization group.

\begin{itemize}
\item
Class 1 ambiguity: Gauge (or equations of motion) ambiguity
\item
Class 2 ambiguity: Scheme ambiguity
\item
Class 3 ambiguity: Local counterterm ambiguity
\end{itemize}

We have already mentioned Class 1 ambiguity at the beginning of this section in order to introduce the concept of gauge invariant flow of coupling constants by $B^I$ functions rather than ambiguous beta functions $\beta^I$ that depends on the gauge we choose. Here, we recapitulate Class 1 ambiguities in more detail.
Due to invariance under the compensating gauge transformation for the coupling constants, the Schwinger functional $W[\gamma_{\mu\nu}, g^I, a_{\mu}]$ is constructed so that it is invariant under the gauge transformation
\begin{align}
\Delta_{w} W[\gamma_{\mu\nu}, g^I,a_\mu] = \int d^3x \sqrt{|\gamma|} \left(D_\mu w \cdot \frac{\delta}{\delta a_{\mu}} - (wg)^I \frac{\delta}{\delta g^I} \right)  W[\gamma_{\mu\nu}, g^I,a_\mu]  = 0 \ \label{gaugeW}
\end{align}
and correspondingly, the form of the Weyl anomaly is ambiguous up to the terms that vanish by \eqref{gaugeW}. In the trace identity, we have seen that the gauge transformation is related to the use of the operator identity
\begin{align}
w \cdot D_\mu J^\mu = -(wg)^I O_I \ .
\end{align}
We call it gauge ambiguity because it is  the gauge transformation on the space-time dependent source terms. In \cite{Nakayama:2012sn}, it was discussed that it corresponds to a certain gauge transformation in $d+1$ dimensional space-time in holography. As we mentioned before, this gauge freedom causes the ambiguities in the definition of beta functions because the choice of the gauge affects the evolution of the scalar coupling constants $g^I$.
This ambiguity in defining beta functions in flat space-time is cancelled if we use the gauge invariant $B^I$ function rather than the $\beta^I$ function in the renormalization group equation \cite{Osborn:1991gm}. Moreover, vanishing of the $B^I$ function is directly related to the Weyl invariance of the theory. In this paper, we mainly focus on the gauge in which the flow of coupling constants is generated by the $B^I$ function although the physics does not change by the choice of gauge.

Class 2 ambiguity is given by the scheme dependence in the renormalization group. Certainly there is an ambiguity in the parameterization of the coupling constant space, given a ``classical action". The parameterization depends on the renormalization  scheme we choose. A well known example is the reparametrization of the scalar coupling constant $g^I \to \tilde{g}^I(g)$. It induces the general coordinate transformation in coupling constant space. Under such reparametrization, various terms transform in rather obvious manners. For instance, $B^I$ and $\hat{\rho}_I$ transform as a vector and one-form respectively, and the anomaly coefficients $C_{IJK}$, $C_{K}$ transform as three-form and one-form. The consistency conditions are manifestly covariant under the reparametrization.\footnote{The situation was a little bit more non-trivial in $d=1+3$ dimension in which some anomaly coefficients do not naturally transform as tensors without further modifications of their definitions \cite{Osborn:1991gm}. We will encounter a similar situation in $d=1+2$ dimension once we introduce the dimensionful coupling constants.}

In a more abstract way, we can generate the scheme ambiguity by considering the variation
\begin{align}
\delta \Delta_{\sigma} &= [\mathcal{D}, \Delta_{\sigma}] \cr
\delta A_{\sigma} & = \mathcal{D} A_{\sigma} 
\end{align}
with any local functional differential operator $\mathcal{D}$ \cite{JO}.\footnote{Practically, we restrict ourselves in the situation where $\mathcal{D}$ preserves the power-counting and the manifest symmetry group $\mathcal{G}$.}
 The above scalar coupling constant reparametrization is generated by choosing
\begin{align}
\mathcal{D} = \int d^3x \sqrt{|\gamma|} f^I(g) \frac{\delta}{\delta g^I} \ , \label{scalarred}
\end{align}
where $\tilde{g}^I = g^I + f^I(g)$ infinitesimally.

A more non-trivial ambiguity in this class is given by the mixing between $a_{\mu}$ and $D_{\mu} g^I$. Choosing 
\begin{align}
\mathcal{D} = \int d^3x \sqrt{|\gamma|} r_I D_{\mu} g^I \cdot \frac{\delta}{\delta a_{\mu} } \ \label{vectorred} 
\end{align}
introduces among other things the shift of the total derivative terms in the trace identity by the amount $\delta v = r_I B^I$. This shift forces us to departure from the original gauge we choose (i.e. $v=0$), so after eliminating this extra $v$ again by Class 1 ambiguity (gauge ambiguity), we can go back to the original gauge with the new parameterization of the local renormalization group:
\begin{align}
\delta \hat{\rho}_I &= (r_I g)^J \hat{\rho}_J - (\hat{\rho}_Ig)^J r_J + (\partial_I r_J - \partial_J r_I)B^J \cr
\delta B^I &= -B^J (r_Jg)^I \  \label{class21}
\end{align}
for the trace identity 
and
\begin{align}
\delta C_{IJK} &= 3C_{L[JK} (r_{I]} g)^L + 2(\partial_{[I}r_{J})C_{K]}  \ \cr
\delta C_I &= C_I (r_Kg)^K + C_K (r_I g)^K  \label{class22}
\end{align}
for the trace anomaly. 

A similar, but a different choice 
\begin{align}
\mathcal{D} = \int d^3x \sqrt{|\gamma|} D_{\mu} w \cdot \frac{\delta}{\delta a_{\mu} } \ 
\end{align}
would just induce the gauge transformation for the background field $a_\mu$, so we could compensate it by transforming the coupling constants $g^I$ using Class 1 ambiguity or the gauge equivalence \eqref{gaugeW}, which leads to a particular choice of the reparametrization of the coupling constants $g^I$ discussed above.

We should note that these ambiguities are all compatible with the consistency conditions proposed in section \ref{sec2.1}.  
At this point, probably it is also worthwhile mentioning that the condition for the conformal invariance $B^I = 0$ in the flat space-time limit with constant source terms is not affected by Class 2 ambiguities.

Finally, Class 3 ambiguity is concerned with the ambiguity in the trace anomaly itself. It is customary that any anomaly is defined only up to local counterterms because we can always add them by hand in the definition of the Schwinger functional. The Schwinger functional is a generating functional for correlation functions of local operators, and the local counterterms do not change the correlation functions except at coincident points in the flat space-time limit, so we can declare that they are arbitrary as long as there are no other constraints from symmetries. Thus we can generate a class of ambiguities in local renormalization group by adding any local functional of coupling constants to the Schwinger functional.

In our discussions of the Weyl anomaly, Class 3 ambiguity therefore shows that the anomalous Weyl variation is arbitrary up to the terms generated by the local counterterms:
\begin{align}
\delta A_{\sigma} = \Delta_{\sigma} W_{\mathrm{local}}[\gamma_{\mu\nu}, g^I, a_{\mu}] \ .
\end{align}
Without any mass parameters, the power-counting demands that the allowed local counterterms are given by
\begin{align}
 W_{\mathrm{local}}[\gamma_{\mu\nu}, g^I, a_{\mu}]  = \int d^3 x \sqrt{|\gamma|} \left( \epsilon^{\mu\nu\rho}  c_{IJK} D_{\mu} g^I D_{\nu} g^J D_{\rho} g^K + \epsilon^{\mu\nu\rho}  f_{\mu\nu} \cdot c_I  \cdot D_{\rho} g^I \right) \ . \label{masslessct}
\end{align} 
As before totally antisymmetric $c_{IJK}$ maps $(R_I \otimes R_J \otimes R_K) \to \mathbf{R}$, and $c_I$ maps $(\mathrm{adj} \otimes R_I) \to \mathbf{R}$ under the compensated symmetry group $\mathcal{G}$.
After some computation, the local counterterms give the ambiguity in the trace anomaly as
\begin{align}
\delta C_{IJK} &= 4B^L\partial_{[L} c_{IJK]} + 3c_{L[JK}(\hat{\rho}_{I]} g)^L +2(\hat{\rho}_{[I} \partial_Jc_{K]}) \cr
\delta C_I &= -3c_{KJI}B^K g^J + B^K(\partial_K c_I -\partial_I c_K)\ . \label{class3trace}
\end{align}

There is a further possible local counterterm given by Chern-Simons terms for the background field $a_{\mu}$:
\begin{align}
W_{\mathrm{local}}[\gamma_{\mu\nu}, g^I, a_{\mu}] = \frac{k_{cs}}{4\pi} \int d^3x \sqrt{|\gamma|} \epsilon^{\mu\nu\rho} \mathrm{Tr} \left(\partial_\mu a_\nu a_{\rho} - \frac{2}{3}a_\mu a_\nu a_\rho \right) \ .
\end{align}
The induced ambiguity in the trace anomaly is
\begin{align}
\delta C_I = \frac{k_{cs}}{4\pi} \hat{\rho}_I \ . \label{cssmb}
\end{align}
Furthermore we could have added the gravitational Chern-Simons term to the Schwinger functional 
as a local counterterm, but it would not contribute to the trace anomaly we are interested in.
The importance of Chern-Simons local counterterms in $1+2$ dimensional quantum field theories has been discussed in the literature \cite{Maldacena:2011nz}\cite{Closset:2012vg}\cite{Closset:2012vp}. Once $k_{cs}$ is quantized from the requirement of the invariance under the large gauge transformation, the ambiguity we discuss here is also quantized.
Since Class 3 ambiguities are generated by the variation of the local functional, it is trivial to see that they satisfy the consistency conditions discussed in section \ref{sec2.1}.

\section{Local renormalization group and consistency conditions in most general cases} \label{sec3}
In this section, we consider the most general forms of the local renormalization group in $d=1+2$ dimension within the power-counting renormalization scheme by adding dimensionful coupling constants to the massless case discussed in section \ref{sec2}.\footnote{Since it does not introduce any interesting new aspects, in this section we will not consider the renormalization of the cosmological constant, which is the source of the identity operator. We present further details on the cosmological constant in appendix A.}
Since the lower dimensional operators (with no additional derivatives) do not mix with the higher dimensional operators in power-counting renormalization scheme, the inclusion of the  dimensionful coupling constants do not alter the massless renormalization group flow in the perturbative search for the conformal fixed point. However, the following discussions may be important in understanding the effect of the composite operator renormalization such as the energy-momentum tensor and mass operators even within the massless renormalization group flow, which have some practical applications such as conformal sequestering and conformal technicolor models.

We introduce the additional ``mass terms" $m^{\alpha} O^{(m)}_{\alpha}$ with mass dimension 2 (e.g. fermion mass terms or scalar quartic interactions) and $M^i O^{(M)}_i$ with mass dimension 1 (e.g. scalar mass terms). Local renormalization group demands that the sources $m^{\alpha}$ and $M^i$ must be space-time dependent.
We suppress the indices $\alpha$ and $i$, which are in certain representations of compensated symmetry group $\mathcal{G}$, in the following to make the notation lighter. The local renormalization group operator is modified with additional terms
\begin{align}
\Delta_{\sigma,m} = - \int d^3 x \sqrt{|\gamma|} \sigma (1-\gamma_{(m)}) m \cdot \frac{\delta}{\delta m} \ 
\end{align}
and
\begin{align}
\Delta_{\sigma,M} &= -\int d^3x \sqrt{|\gamma|}\left(\sigma(2-\gamma_{(M)}) M + \frac{1}{4} \sigma R \eta + \sigma \delta_I (D^2 g^I) + \sigma \epsilon_{IJ} (D^\mu g^I D_\mu g^J) \right. \ \cr
 & \left. \left. + 2\partial_\mu \sigma (\theta_I D^\mu g^I) + (D^2 \sigma) \tau + \sigma m \cdot \kappa \cdot m \right) \cdot \frac{\delta}{\delta M} \right) \ \label{mderiv}
\end{align}
from the simple power-counting.
Hereafter $\cdot$ implies the summation over $\alpha$ and $i$ induced by the inner product of the symmetry group. With these additional contributions, the total local renormalization group operator is now modified as
\begin{align}
\Delta_{\sigma} =&  \int d^3x \sqrt{|\gamma|} \left( 2\sigma \gamma_{\mu\nu} \frac{\delta}{\delta \gamma_{\mu\nu}} + \sigma \beta^I \frac{\delta}{\delta g^I}  \right. + \left. \left( \sigma \rho_I D_{\mu} g^I - (\partial_\mu \sigma) v \right)\cdot \frac{\delta}{\delta a_{\mu}} \right) \cr
& + \Delta_{\sigma,m} + \Delta_{\sigma,M} \ . 
\end{align}

They correspond to the additional terms in the trace identity
\begin{align}
T^\mu_{\ \mu}|_{M,m} &= (\gamma_{(m)}-1) m \cdot O^{(m)} + (\gamma_{(M)}-2) M \cdot O^{(M)} -  \frac{1}{4} R \eta \cdot O^{(M)} - D^2 (\tau \cdot O^{(M)}) \cr
& - \delta_I (D^2 g^I) \cdot O^{(M)} - \epsilon_{IJ} (D^\mu g^I D_\mu g^J ) \cdot O^{(M)} \cr
& + 2 D_{\mu} (\theta_I D^{\mu} g^I \cdot O^{(M)}) - m \cdot \kappa \cdot m \cdot O^{(M)} \ 
\end{align}
with the Schwinger action principle:
\begin{align}
\frac{\delta}{\delta m(x)} W &= - \langle O^{(m)} (x) \rangle \cr 
\frac{\delta}{\delta M(x)}W &= - \langle O^{(M)}(x) \rangle \ .
\end{align}

At this point, it is instructive to understand the meaning of some coefficients in the trace identity as the operator mixing under the massless renormalization group. From the local renormalization group equation combined with the power-counting, we obtain the operator mixing in the flat space-time limit with constant coupling constants  \cite{Osborn:1991gm}:
\begin{align}
\frac{d}{d\log\mu}\left(\begin{array}{c}
T^{\mu}_{\ \mu}  \\
O^{(M)} \\
O_I
\end{array}
\right)
= \left(
\begin{array}{ccc}
0  & \eta \Box & 0 \\
0 & -\gamma_{(M)} & 0 \\
0 & \delta_I \Box & -\gamma_I^{\ J} \\
\end{array}
\right) \left(
\begin{array}{c}
T^{\mu}_{\ \mu}  \\
O^{(M)} \\
O_J
\end{array}
\right) . \label{compr}
\end{align}
Here $\gamma_{(M)}$ is interpreted as the mass anomalous dimension matrix for operators $O^{(M)}$, and $\gamma_I^{\ J} = \partial_I B^J + (\hat{\rho}_Ig)^J$ as the anomalous dimension matrix for dimension 3 scalar operators $O_I$.\footnote{The gauge rotation by $\hat{\rho}_I$ is necessary from Class 1 ambiguity. The combination is what appears in the modified Lie derivative \eqref{modlie} introduced in  \cite{Osborn:1991gm}\cite{JO}, and we will see how this gives the expected result in supersymmetric field theories in section \ref{sec4.2}.} 
Similarly, $\delta_I$ terms are interpreted as the mixing between $O_I$ and $\Box O^{(M)}$ under renormalization.
We will see that the renormalization of the curvature coupling term $\eta$ 
can be related to the  other terms as a consequence of the consistency conditions. Physically, this $\eta$ term is the main source of the renormalization of the energy-momentum tensor as 
\begin{align}
\frac{d}{d\log\mu} T_{\mu\nu} =-\frac{1}{2}(\partial_\mu \partial_\nu - \Box \eta_{\mu\nu})  \eta O^{(M)} \label{renormalem}
\end{align}
 and it may play an important role in cosmology. Note that the right hand side is automatically conserved irrespective of the nature of $O^{(M)}$, and it is consistent with the conservation of the renormalized energy-momentum tensor at every energy scale. We also note that the global energy and momenta are not renormalized despite the renormalization of the energy-momentum tensor.

With the presence of the dimensionful coupling constants, the anomalous Weyl variation of the Schwinger functional acquires new terms
\begin{align}
& A_{\sigma;M,m} = \cr
&  \int d^3x \sqrt{|\gamma|}\left( \sigma( M \cdot \beta \cdot m - \frac{1}{4} R I \cdot m  + J_I(D^2 g^I) \cdot m + K_{IJ} (D_\mu g^I D^\mu g^J) \cdot m + Sm^3) \right. \cr
 & \left. - 2\partial_\mu \sigma (L_I D^\mu g^I \cdot m) + (D^2 \sigma) k\cdot m \right) \ , \label{massanomaou}
\end{align}
where we assume $K_{IJ} = K_{(IJ)}$ is symmetric, and $Sm^3$ is a shorthand notation for $S_{\alpha\beta\gamma} m^{\alpha}m^{\beta}m^{\gamma}$.
They correspond to the additional terms in the trace anomaly
\begin{align}
A_{\mathrm{anomaly};M,m} & =  -M \cdot \beta \cdot m + \frac{1}{4} R I \cdot m - J_I (D^2 g^I)  \cdot m - K_{IJ} (D_\mu g^I D^\mu g^J) \cdot m -  Sm^3 \cr
 & -2D_{\mu} (L_I D^\mu g^I \cdot m) - D^2 (k\cdot m) \ . 
\end{align}
Note that unlike the situation in section \ref{sec2}, the trace anomaly may not vanish even in the flat space-time limit with constant sources. This is because the power-counting allows that the cosmological constant is renormalized when the mass parameters are present. At the conformal fixed point, some of these terms are computed in \cite{Bzowski:2013sza}.

\subsection{Consistency condition}\label{sec3.1}
We can repeat the same analysis for the consistency conditions of local renormalization group with additional mass parameters. As discussed in section \ref{sec2.1}, there are two distinct classes of consistency conditions from the integrability condition $[\Delta_{\sigma}, \Delta_{\tilde{\sigma}}] = 0$ of the local renormalization group operator.

Class 1 consistency condition (Integrability conditions for the local renormalization group transformation operator) is obtained by requiring $[\Delta_{\sigma}, \Delta_{\tilde{\sigma}}] = 0$ as a differential operator acting on arbitrary functional $W[\gamma_{\mu\nu}, g^I, a_{\mu}, m, M]$. With the additional dimensionful parameters, in addition to the previous constraint \eqref{rhocons}, we must require (see appendix \ref{appb} for Weyl variations) 
\begin{align}
\eta  &= \delta_I B^I - (B^I \partial_I \tau - \gamma_{(M)}) \tau \cr
\delta_I + 2(\partial_I B^J + \frac{1}{2}(\hat{\rho}_I g)^J) \delta_J + 2 \epsilon_{IJ} B^J &= 2(\tilde{\mathcal{L}}_{B,\hat{\rho}} - \gamma_{(M)}) \theta_I \ .  \label{consis}
\end{align}
Here the modified Lie derivative \cite{Osborn:1991gm}\cite{JO} 
\begin{align}
\tilde{\mathcal{L}}_{B,\hat{\rho}} \theta_I = B^J\partial_J \theta_I + (\partial_I B^J+(\hat{\rho}_I g)^J) \theta_J = B^J \partial_J \theta_I + \gamma_{I}^{\ J} \theta_J \label{modlie}
\end{align}
for the 1-form is introduced (we will use the similar definition for the other tensors).
Note that the first equation \eqref{consis} determines $\eta$ from the other parameters in the trace anomaly.\footnote{The $\eta$ term in the trace anomaly is a genuine geometric obstruction for the Weyl transformation in $(1+2)$ dimension, but as we will discuss, we can make it vanish at conformal fixed point by choosing the judicious counterterms.} The necessity of the first equation can be also seen from the consistency of the trace identity
\begin{align}
T^{\mu}_{\ \mu} = B^I O_I - \tau \cdot \Box O^{(M)} \ . 
\end{align}
 under the massless renormalization group with the composite operator renormalization \eqref{compr} in the flat space-time limit with constant sources. 

We emphasize again that Class 1 consistency condition is rather universal and the structure is not very much different from the one that appeared in $d=1+3$ \cite{Osborn:1991gm}\cite{JO} with mass parameters. We can also understand the universality from the above argument that the consistency condition is a consequence of the trace identity and the composite operator renormalization.

Instead, Class 2 consistency conditions  (Integrability conditions for the Weyl anomaly) deal with the anomalous variation and the subsequent conditions will be unique to $d=1+2$ dimension. By demanding
\begin{align}
\Delta_{\tilde{\sigma}} A_{\sigma} = \Delta_{\sigma} A_{\tilde{\sigma}} \ 
\end{align}
in the new terms in Weyl anomaly \eqref{massanomaou}, we obtain the new constraint:
\begin{align}
I + B^I J_I - \tau \beta &= B^I \partial_I k + \gamma_{(m)} k \ \cr
\frac{1}{2} J_I + (\partial_I B^J + \frac{1}{2}(\hat{\rho}_I g)^J )J_J + K_{IJ} B^J + \tilde{\mathcal{L}}_{B,\hat{\rho}} L_ I + \gamma_{(m)} L_I &= \theta_I \beta \ 
\label{massconsist}
\end{align}
in addition to  \eqref{consistenccc}.
Unlike in $d=1+3$ discussed in \cite{Osborn:1991gm}\cite{JO}, the consistency conditions \eqref{consistenccc} for the beta functions for dimensionless coupling constants are not modified by the presence of the dimensionful coupling constants.

\subsection{Ambiguity} \label{sec3.2}
The ambiguities in massless renormalization group discussed in section \ref{sec2.2} can be extended to the most generic renormalization group with the dimensionful parameters. We have three distinct classes of ambiguities. 

Class 1 ambiguities (Gauge ambiguity) appear due to the gauge invariance of the Schwinger functional $W[\gamma_{\mu\nu}, g^I, a_{\mu}, m, M]$. 
The gauge invariance must be extended to include the dimensionful operators:
\begin{align}
\Delta_{w} = \int d^3x \sqrt{|\gamma|} \left(D_\mu w \cdot \frac{\delta}{\delta a_{\mu}} - (wg)^I \frac{\delta}{\delta g^I} - (wM) \cdot \frac{\delta}{\delta M} - (wm) \cdot \frac{\delta}{\delta m} \right) = 0 \ ,
\end{align}
which corresponds to the operator identity
\begin{align}
w \cdot D_\mu J^\mu = -(w g)^I O_I - (wM) \cdot O^{(M)} - (wm) \cdot O^{(m)} \ . 
\end{align}
By using this ambiguity, we can always remove the total derivative term $D_\mu (v\cdot J^\mu)$ in the trace identity with $\beta^I \to B^I = \beta^I - (vg)^I$ and so on.\footnote{
In principle this equation could contain additional terms $(w \alpha_R) R O^M + (w \alpha_d) D^2 O^M$.} In section \ref{sec3.1}, it was assumed that this gauge ambiguity is fixed by requiring there is no $w\cdot D_{\mu} J^\mu$ term in the trace anomaly. This is the most convenient gauge choice because vanishing of $B^I$ function together with vanishing of dimensionful parameters (e.g. $M$ and $m$) will imply the Weyl invariance of the theory up on the improvement of the energy-momentum tensor that we will discuss in a moment.

Class 2 ambiguities (Scheme ambiguity) are related to the scheme choice of the local renormalization group. The simplest example is the reparametrization $g^I \to \tilde{g}^I(g^J)$ of the dimensionless scalar coupling constants, which is usually associated with the choice of the renormalization schemes. Most of the consistency equations are manifestly covariant under such reparametrization, but some consistency equations (e.g. second lines of \eqref{consis} and \eqref{massconsist}) are not manifestly covariant because ordinary derivatives with respect to $I$ appears rather than covariant derivatives or Lie derivatives. However, some coefficients such as $\epsilon_{IJ}$ and $K_{IJ}$ transforms non-covariantly due to $D^2 g^I$ terms in \eqref{mderiv} and \eqref{massanomaou} so that the consistency conditions are actually covariant as they should be. 

More generally, we can generate the scheme ambiguity by considering the variation
\begin{align}
\delta \Delta_{\sigma} &= [\mathcal{D}, \Delta_{\sigma}] \cr
\delta A_{\sigma} & = \mathcal{D} A_{\sigma} 
\end{align}
with any local functional differential operator $\mathcal{D}$. 
The above mentioned reparametrization ambiguity is induced by
\begin{align}
\mathcal{D} = \int d^3 x   \sqrt{|\gamma|} \left( f_{g}^I \frac{\delta}{\delta g^I} + f_m m \cdot \frac{\delta}{\delta m} + (f_M M + m f_{Mm} m) \cdot \frac{\delta}{\delta M}\right) \ .
\end{align}
We have included the additionally possible reparametrization of mass parameters $\delta m = f_{m} m$ and $\delta M = f_M M + m f_{Mm} m$. 
In addition, we have other Class 2 ambiguities for the mixing between $D_\mu g^I$ and $a_\mu$ as
\begin{align}
\mathcal{D} = \int d^3x \sqrt{|\gamma|} r_I D_\mu g^I \frac{\delta}{\delta a_\mu} \ ,
\end{align}
which, in addition to \eqref{class21} we have already obtained in the massless case, induces
\begin{align}
\delta\delta_I &= (r_Ig)^J \delta_J \cr
\delta \theta_I &= (r_I g)^J \theta_J \cr
\delta \epsilon_{IJ} &= (r_Ig)^K \epsilon_{KJ} + (r_Jg)^K \epsilon_{IK} + (\partial_{(I} r_{J)} g )^K \delta_K + 
2\delta_{K}(r_{(I})^K_{\ J)} \ .
\end{align}
In the last line, explicit matrix notation of $(r_{I})^{K}_{\ J} = r_{I}^a (T_a)^{K}_{\ J}$ is used.
At the same time, the trace anomaly is modified, in addition to \eqref{class22}, as
\begin{align}
\delta K_{IJ} &= (r_Ig)^K K_{KJ} + (r_Jg)^K K_{IK}  +(\partial_{(I}r_{J)}g)^KJ_K + 2J_{K}(r_{(I})^K_{\ J)} \cr
\delta L_I &= (r_Ig)^J L_J \cr
\delta J_I &=  (r_Ig)^J J_J  \  . 
\end{align}

Furthermore, we have extra Class 2 ambiguity for the mixing between $R$, $D^2 g^I$ and $D_\mu g^I D^\mu g^J$ with
\begin{align}
\mathcal{D} = \int d^3x \sqrt{|\gamma|} \left(\frac{1}{4}R h + (D^2 g^I) d_I + (D_\mu g^I D^\mu g^J) e_{IJ} \right) \cdot \frac{\delta}{\delta M} \ ,
\end{align}
where we assume $e_{IJ} = e_{(IJ)}$ is symmetric.
Under this scheme change associated with the field redefinition, we obtain
\begin{align}
\delta \eta &= (B^I \partial_I h - \gamma_{(M)} h) \cr
\delta \tau &= -h + d_I B^I \cr
\delta \theta_I &= \frac{1}{2}d_I + \left(\partial_IB^J + \frac{1}{2}(\hat{\rho}_Ig)^J \right)d_J + e_{IJ} B^J \cr
\delta \delta_I &= (\tilde{\mathcal{L}}_{B,\hat{\rho}} -\gamma_{(M)}) d_I \cr
\delta \epsilon_{IJ} &= (\tilde{\mathcal{L}}_{B\,\hat{\rho}} - \gamma_{(M)}) e_{IJ} +
(\partial_I \partial_J B^K + (\partial_{(I}(\hat{\rho}_{J)})g)^K)d_K + 2d_K(\hat{\rho}_{(I})^K_{J)} \ 
\end{align}
as well as the change in the trace anomaly
\begin{align}
\delta I &= -4h\beta \cr
\delta J_I &=  \beta d_I \cr
\delta K_{IJ} &=  \beta e_{IJ} \ . 
\end{align}
In particular, one may always set $\tau = \theta_I = 0$ by using the ambiguity. We note that $\tau = 0$ choice is nothing but the improvement of the energy-momentum tensor so that the $\Box O^{(M)}$ term is absent in the trace anomaly in the flat space-time as we will discuss shortly.

Finally Class 3 ambiguities (Counterterm ambiguity) are induced by the local counterterms in the Schwinger functional. With the presence of the dimensionful coupling constants, the new local counterterms we could add in addition to \eqref{masslessct} are
\begin{align}
\int d^3x \sqrt{|\gamma|} \left( M \cdot \mathcal{B}_{Mm} \cdot m -\frac{1}{4} R \mathcal{I} \cdot m  + \mathcal{J}_I (D^2 g^I)  \cdot m +  \mathcal{K}_{IJ}  (D_\mu g^I D^\mu g^J)\cdot m + \mathcal{S} m^3 \right) \ , \label{class3massive}
\end{align}
where we assume $\mathcal{K}_{IJ} = \mathcal{K}_{(IJ)}$ is symmetric, and $\mathcal{S} m^3$ is a shorthand notation for $\mathcal{S}_{\alpha\beta\gamma} m^{\alpha}m^{\beta}m^{\gamma}$
These counterterms induce the modification of the trace anomaly as
\begin{align}
\delta \beta &= B^K\partial_K\mathcal{B}_{Mm} + \gamma_{(M)} \mathcal{B}_{Mm} + \mathcal{B}_{Mm} \gamma_{(m)} \cr
\delta I &= \eta \mathcal{B}_{Mm} + \mathcal{I} \gamma_{(m)} + B^K \partial_K \mathcal{I} \cr
\delta J_I  &= -\delta_I \mathcal{B}_{Mm} +\tilde{\mathcal{L}}_{B,\hat{\rho}}\mathcal{J}_I +\mathcal{J}_I \gamma_{(m)} \cr
\delta K_{IJ} &= \tilde{\mathcal{L}}_{B,\hat{\rho}} \mathcal{K}_{IJ} + \mathcal{K}_{IJ}\gamma_{(m)}- \epsilon_{IJ} \mathcal{B}_{Mm} +(\partial_I \partial_JB^K +(\partial_{(I}(\hat{\rho}_{J)})g)^K) \mathcal{J}_K +2\mathcal{J}_K(\hat{\rho}_{(I})^K_{J)}  \cr
\delta S & = -\kappa \mathcal{B}_{Mm} +B^K\partial_K \mathcal{S} + 3\gamma_{(m)} \mathcal{S} \cr
\delta L_I &= \theta_I \mathcal{B}_{Mm} -\frac{1}{2}\mathcal{J}_I - (\partial_IB^K) \mathcal{J}_K - \frac{1}{2}(\hat{\rho}_I g)^K \mathcal{J}_K - \mathcal{K}_{IJ} B^J \cr
\delta k &= -\tau \mathcal{B}_{Mm} - \mathcal{I} + B^I \mathcal{J}_I \label{3amb} ,
\end{align}
where $3\gamma_{(m)} \mathcal{S}$ really means $(\gamma_{(m)}^{\alpha\alpha'} + \gamma_{(m)}^{\beta\beta'} + \gamma_{(m)}^{\gamma \gamma'}) \mathcal{S}_{\alpha' \beta' \gamma'}$. With these ambiguities, we may set $k= L_I=0$.

To conclude this section, let us address some applications of the local renormalization group with mass parameters. In particular, we address some properties of the energy-momentum tensor under renormalization.

The first application is concerned with how to construct the renormalization group invariant energy-momentum tensor. 
For many applications, it is important to understand the renormalization of the energy-momentum tensor and possible improvements. Generally, the energy-momentum tensor in flat space-time is ambiguous under the improvement
\begin{align}
T_{\mu\nu} \to T_{\mu\nu} + (\partial_\mu\partial_\nu - \Box \eta_{\mu\nu}) L
\end{align}
for any scalar operator $L$.\footnote{More generically spin 2 (or higher) improvement is possible \cite{Polchinski:1987dy} (in particular in non-unitary theories) but it is not relevant for our discussions.} In the local renormalization group with curved space-time background, we have already argued that by using Class 3 ambiguity induced by $h$, we can always set $\tau = 0$. This convention is know as the Callan-Coleman-Jackiw improved energy-momentum tensor \cite{Callan:1970ze}. One advantage of the choice is that when $B^I = 0$ at the fixed point, the theory is manifestly conformal invariant in the flat space-time and we keep the same property during the renormalization by adjusting $h$ at each energy scale. Actually, Class 2 consistency condition \eqref{consis} tells that it is even Weyl invariant in the curved background when $M = m = 0$ with constant coupling constants at the fixed point  because the curvature term in the trace anomaly also vanishes $\eta=0$. This is the energy-momentum tensor implicitly assumed in \cite{Komargodski:2011vj}.

However, away from the conformal fixed point, this improved energy-momentum tensor may be renormalized according to \eqref{renormalem} due to the operator mixing.  Indeed, Class 2 consistency condition \eqref{consis}  tells that this is unavoidable as long as $\delta^I \neq 0$. 
For this reason, it may be sometimes more useful to define the non-renormalized energy-momentum tensor by demanding $\eta = 0$ rather than $\tau = 0$. This is known as Zamolodchikov's canonically scaling energy-momentum tensor \cite{Zamolodchikov:1986gt}\cite{Polchinski:1987dy} (see also \cite{Yonekura:2012kb}). As argued by Polchinski,\footnote{There is a typo in eq (18) of \cite{Polchinski:1987dy}. We would like to thank Z.~Komargodski for the related discussion.} this is always possible by adjusting $h$ when $\gamma_{(M)}$ does not contain any zero eigenvalues, being invertible. Otherwise, due to a potential obstruction to choose $\eta=0$, it is logically possible that the theory is scale invariant, but the energy-momentum tensor is still logarithmically renormalized. 
 When the theory is conformal invariant (i.e. $B^I = 0$) then such a possibility is unavailable from the consistency conditions (e.g. \eqref{consis}). In any case, away from the fixed point, it is important to understand that the Callan-Coleman-Jackiw improved energy-momentum tensor and Zamolodchikov's non-renormalized energy-momentum tensor (if any) may differ.

Another potentially interesting application of the massive local renormalization group analysis is the renormalization of the Einstein-Hilbert term that appears as $I$ in the trace anomaly. We have already discussed that one can always set $k=0$ by using Class 3 ambiguity. If we further use the  Callan-Coleman-Jackiw improved energy-momentum tensor (i.e. $\tau = 0$), we see that the Einstein-Hilbert term is not renormalized at the conformal fixed point $B^I = 0$. Alternatively, by using non-zero $k$, we may be able to set $I=0$ and try to keep the non-renormalization of the Einstein-Hilbert term away from the fixed point whenever $\gamma_{(m)}$ does not contain any zero eigenvalues. Needless to say, regardless of the possibility to obtain the non-renormalized Einstein-Hilbert term discussed here, the actual value of the Einstein-Hilbert term can be changed in an arbitrary manner (at a given renormalization scale) by adding the local counterterm.

\section{Checks of consistency conditions}\label{sec4}
So far, our discussions have been rather formal. 
In this section, we would like to perform modest checks of our arguments on the local renormalization group in some examples. Of course, our discussions must apply to perturbation theories based on Feynman diagrams in any renormalization scheme, but we would like to show the generality of our results from the other ways to compute beta functions and the trace anomaly in renormalization group.

\subsection{Conformal perturbation theory}\label{sec4.1}
To begin with, we would like to compute beta functions for vector operators (i.e. $v$ and $\rho_I$) in conformal perturbation theory (see also \cite{Nakayama:2013is}). We note that the conventional perturbation theory based on Feynman diagrams is just an example of conformal perturbation theory around a free (massless ultraviolet) fixed point.
Here we start with a general conformal field theory and perturb it by adding marginal scalar interactions 
$\delta S = \int d^3x g^I(x) O_I(x)$. In order to facilitate the computation of the vector beta functions, we have introduced the space-time dependent coupling constants $g^I(x)$. At the order we are interested in, the curvature of the space-time is not important.

We assume that the scalar operator $O_I(x)$ has the canonical normalization
\begin{align}
\langle O_I(x) O_J(y) \rangle_0  = \frac{\delta_{IJ}}{(x-y)^6} \ 
\end{align}
in the reference conformal field theory at the ultraviolet fixed point. For simplicity we have assumed that the operators $O_I(x)$ are conformal primaries with dimension $\Delta_I = 3$ in the reference conformal field theory, but generalizations to a slightly relevant perturbation are possible (within the so-called Zamolodchikov scheme \cite{Zamolodchikov:1987ti}).

In order to compute the scalar beta functions as well as vector beta functions, we assume the operator product expansion:
\begin{align}
O_I(x) O_J(y) = \frac{\mathcal{C}_{IJK}}{(x-y)^3}O_K(y) + \frac{\mathcal{C}^{a}_{IJ}(x-y)_\mu}{(x-y)^{5}} J_{a}^\mu (y) + \cdots , \label{ope}
\end{align}
where the operator product expansion coefficient $\mathcal{C}_{IJK}$ is totally symmetric and $\mathcal{C}^a_{IJ} = -\mathcal{C}^a_{JI}$ is a certain representation matrix of the flavor symmetry group (denoted by $\mathcal{G}$ before) generated by $J_a^\mu$. In the reference conformal field theory, the current $J_a^{\mu}$ is conserved with conformal dimension $\Delta_a = 2$. The appearance of $\mathcal{C}^{a}_{IJ}$ in the scalar operator product expansion means that the current conservation is violated by the perturbation \cite{Friedan:2012hi}\cite{Nakayama:2013is} as
\begin{align}
\partial_\mu J^{\mu}_a = g^I \mathcal{C}_{IJ}^a O^J  \ . 
\end{align}

At the second order in conformal perturbation theory, we have to evaluate and renormalize the divergent integral in the evaluation of the Schwinger functional
\begin{align}
\delta W = \left\langle \int d^3x d^3y g^I(x) O_I(x) g^J(y) O_J(y) \right \rangle_{0} 
\end{align}
by using the above operator product expansion. The scalar part of the operator product expansion gives a diverging factor
\begin{align}
\delta W|_{\mathrm{scalar}}  \sim \left\langle 2\pi \int d^3z \log\mu \mathcal{C}_{IJK} g^I(z) g^J(z) O_K(z) \right \rangle_{0} \ ,
\end{align}
which gives the scalar beta function
\begin{align}
\beta^I = \frac{dg^I}{d\log\mu} = 2\pi \mathcal{C}_{IJK} g^J g^K + \mathcal{O}(g^3) \ .
\end{align}
Similarly, from the current part of the operator product expansion gives another diverging contribution 
\begin{align}
\delta W|_{\mathrm{vector}}  \sim \left\langle 2\pi \int d^3z \log\mu g^I(z) \partial_\mu g^J(z) \mathcal{C}_{IJ}^a J^\mu_a(z) \right \rangle_0 \ ,
\end{align}
which results in the renormalization of the background gauge fields $a_\mu$ with
\begin{align}
\rho^a_I &= 2\pi \mathcal{C}_{IJ}^a g^J \cr
 v&= 0 \ .
\end{align}
It is possible to change the renormalization prescription so that $v$ is non-zero by using the equations of motion or gauge transformation of the background source fields \cite{Nakayama:2013is}, but it does not affect the following argument because we work on the gauge invariant $B^I$ functions and $\hat{\rho}_I$ functions.

At the second order in conformal perturbation theory, we therefore conclude
\begin{align}
B^I &= 2\pi \mathcal{C}_{IJK} g^J g^K \cr
\hat{\rho}_I^a &= 2\pi \mathcal{C}_{IJ}^a g^J \ .
\end{align}
As a check of our formal argument in section \ref{sec2}, we immediately realize
\begin{align}
B^I \hat{\rho}_I^a = 0 \ 
\end{align}
due to the symmetry of $\mathcal{C}_{IJK}$ and anti-symmetry of $\mathcal{C}_{IJ}^a$. Thus, the transversality condition is satisfied. At a higher order, this becomes more non-trivial because apparently the computation of $B^I$ and $\hat{\rho}_I$ are not immediately related with each other in particular at different orders in perturbations theory (see however the supersymmetric case in section \ref{sec4.2}).

We have a couple of technical remarks about the above computation. 

\begin{itemize}
\item

In the above evaluation of the divergent integral, we had to keep track of (the absence of) the total derivative terms. We used the Polyakov regularization \cite{Polyakov:1981re} $\lim_{x\to y} \log(x-y)|_{\mathrm{reg}} = \log \sigma(x)$ in  order to take into account the position dependent cut-off scale. At the second order in conformal perturbation theory, this is the most natural prescription, but at higher orders, it may be more practical to use the dimensional regularization because the total derivative terms will not affect the bare energy-momentum tensor in $d = 3-\epsilon$ dimension, and total derivative terms in counterterms can be discarded safely. A systematic way to compute the higher order vector beta functions in dimensional regularization with minimal subtraction was thoroughly developed in \cite{Jack:1990eb}\cite{JO} (see also \cite{Fortin:2012hn}).

\item
Once we try to evaluate the integral in the dimensional regularization, we have to assign the scaling dimensions of the operators  $O^I$ (called $k^I$ in \cite{Jack:1990eb} as $\Delta_I = 3-k^I \epsilon$) in $3-\epsilon$ dimension. In conventional Lagrangian field theories, these are naturally determined by the wavefunction renormalization of the kinetic operators in $d=3-\epsilon$ dimension, but it is not obvious how it works in general conformal perturbation theory without explicit Lagrangian. However, we can check that this ambiguity cancels out in the final computation of the trace of the energy-momentum tensor because the energy-momentum tensor in $d=3-\epsilon$ dimension also contains the additional contributions that are related to $k^I$ from $\beta^I_{d = 3-\epsilon} = k^I g^I + \beta^I_{d=3}$ and $T^{\mu}_{\ \mu} = \beta^I_{3-\epsilon} O_I + \cdots$ which eventually led to the explicit loop counting factor in the dimensional regularization formula found in \cite{JO} (see also \cite{Fortin:2012hn} for the appearance of $k^I$ in the computation of $v$ there). This cancellation is reassuring because the ``loop counting" is different from the order of conformal perturbation, and the explicit appearing of the former in the computation of vector beta functions seems mysterious from the conformal perturbation theory viewpoint.

\end{itemize}

Let us briefly discuss the trace anomaly induced by the space-time dependent coupling constant within the conformal perturbation theory. In principle, it should be computable by evaluating the vacuum energy in conformal perturbation theory and renormalize it. In order to compute the contribution to the term
\begin{align}
\epsilon^{\mu\nu\rho} C_{IJK}(g) \partial_\mu g^I \partial_\nu g^J \partial_\rho g^K 
\end{align}
in the trace anomaly, for instance, we have to break the CP invariance due to the appearance of $\epsilon^{\mu\nu\rho}$. Such breaking is not encoded in the leading order operator product expansion \eqref{ope} nor in the normalization of the two-point function in a manifest manner.
 In this way, we have to evaluate the vacuum energy at least fourth order in perturbation theory (and probably fifth order to break the CP from the scalar perturbations alone) to obtain non-zero results. Unfortunately, there is no systematic way to evaluate the conformal perturbation theory at that order since we need the full spectrum and operator product expansion to compute the correlation functions, so we would like to defer the actual computation for a future problem.

\subsection{Supersymmetry}\label{sec4.2}
While our discussions so far do not assume supersymmetry, it is possible to check some of our results to all order in perturbation theory if we assume $\mathcal{N}=2$ supersymmetry in $d=1+2$ dimension (we follow the superspace convention of \cite{Dumitrescu:2011iu}). 
Let us consider the  Wess-Zumino model with dimensionless coupling constants
\begin{align}
 W = Y^{abcd} \Phi_a \Phi_b \Phi_c \Phi_d ,
\end{align}
where $\Phi^a$ $(a=1,\cdots N)$ are chiral superfields and the flavor symmetry group $\mathcal{G}$ compatible with $\mathcal{N}=2$ supersymmetry is $U(N)$ (in addition to the $U(1)$ R-symmetry).
In order to discuss the local renormalization group with the manifest supersymmetry, we uplift the coupling constants $Y^{abcd}$ to chiral superfields.
The usual argument based on the holomorphy and R-symmetry tells that the divergence to all orders in perturbation theory can be removed by the counterterm in the K\"ahler potential
\begin{align}
\mathcal{L}_{\mathrm{ct}} = \int d^4 \theta K^{a\bar{b}}(Y,\bar{Y}) \Phi_a \bar{\Phi}_{\bar{b}} \ .
\end{align}

One consequence of the supersymmetric non-renormalization theorem is that the beta function for $Y_{abcd}$ is completely determined from the anomalous dimension matrix
\begin{align}
\beta_{Y^{abcd}} = \gamma^{a \bar{e}} Y^{ebcd} + \gamma^{b\bar{e}}Y^{aecd} + \gamma^{c\bar{e}} Y^{abed} + \gamma^{d\bar{e}} Y^{abce} \ . \label{susybeta}
\end{align}
Here, the anomalous dimension matrix $\gamma^{a\bar{b}}(Y,\bar{Y})$ is obtained from the renormalization of the K\"ahler potential counterterm as
\begin{align}
\gamma^{a\bar{b}} = \frac{d K^{a\bar{b}}}{d \log \mu} \ . 
\end{align}
The unitarity demands that the K\"ahler potential $K^{a\bar{b}}$ hence $\gamma^{a\bar{b}}$ is Hermitian.

On the other hand, the same K\"ahler potential determines the vector beta functions for the $U(N)$ rotations \cite{Nakayama:2012nd}\cite{Fortin:2012hc}\cite{JO}:
\begin{align}
[\rho_{abcd} dY^{abcd} + \bar{\rho}_{\bar{a}\bar{b}\bar{c}\bar{d}} d\bar{Y}^{\bar{a}\bar{b}\bar{c}\bar{d}}]^{e\bar{f}} = - (\partial_{Y^{abcd}} \gamma^{e\bar{f}}) dY^{abcd} +   (\partial_{\bar{Y}^{\bar{a}\bar{b}\bar{c}\bar{d}}} \gamma^{e\bar{f}}) d\bar{Y}^{\bar{a}\bar{b}\bar{c}\bar{d}} 
\end{align}
from the $\bar{\theta} \sigma^\mu \theta$ terms in $K^{a\bar{b}}$.
Assuming that the computation is done in dimensional regularization (in order to avoid the complexity due to total derivatives), the counterterm also determines
\begin{align}
v^{e\bar{f}} = i\frac{\partial \gamma^{e\bar{f}}}{\partial Y^{abcd}}Y^{abcd} - i\frac{\partial \gamma^{e\bar{f}}}{\partial \bar{Y}^{\bar{a}\bar{b}\bar{c}\bar{d}}} \bar{Y}^{\bar{a}\bar{b}\bar{c}\bar{d}} = 0 
\end{align}
 in the holomorphic scheme we use here.

The anomalous dimension of the chiral operators that appear in the superpotential must be determined from $\gamma^{a\bar{b}}$. We can confirm that this is the case by using  $\gamma_I^{\ J} = \partial_I B^J + (\hat{\rho}_Ig)^J$ with the above formula for the beta functions.  Notice that the additional rotation by $\hat{\rho}_I$ is important to cancel various unwanted mixing from $\partial_I B^J$ alone.

In section \ref{sec2.1}, we have shown that Class 1 consistency condition demands that 
\begin{align}
B^I \hat{\rho}_I =  0 \ ,
\end{align}
which is equivalent to
\begin{align}
\frac{\partial \gamma^{e\bar{f}}}{\partial Y^{abcd}}\beta_{{Y^{abcd}}} - \frac{\partial \gamma^{e\bar{f}}}{\partial \bar{Y}^{\bar{a}\bar{b}\bar{c}\bar{d}}} \beta_{{\bar{Y}^{\bar{a}\bar{b}\bar{c}\bar{d}}}} = 0 \ ,  \label{nontrivial}
\end{align}
where $\beta_{{Y^{abcd}}} $ can be expressed by \eqref{susybeta} with the anomalous dimension matrix.

We can see that this condition is true at each order in supergraph computations  of the anomalous dimensions \cite{Jack:1999aj}\cite{JO}. Operationally, what $
\frac{\partial \gamma^{e\bar{f}}}{\partial Y^{abcd}}\beta_{{Y^{abcd}}} $ does is adding extra anomalous dimension factor to each $\Phi_a \to \bar{\Phi}_{\bar{a}}$ lines in supergraph computation of the wavefunction renormalization. Since every propagator is oriented as $\Phi_a \to \bar{\Phi}_{\bar{a}}$ in the computation for wavefunction renormalization (due to R-symmetry), the action of  $\frac{\partial \gamma^{e\bar{f}}}{\partial \bar{Y}^{\bar{a}\bar{b}\bar{c}\bar{d}}} \beta_{{\bar{Y}^{\bar{a}\bar{b}\bar{c}\bar{d}}}}$ does exactly the same thing and \eqref{nontrivial} holds. It would be interesting to see if there is a more direct proof without relying on the supergraph.

In specific to $d=1+2$ dimension, let us discuss the possible $\mathcal{N}=2$ supersymmetric extension of the Weyl anomaly. The Weyl anomaly is replaced by super Weyl anomaly generated by a chiral superfiled $\Sigma$.
We can easily construct the supersymmetric generalization of the Weyl anomaly terms. For instance, if the symmetry group $\mathcal{G}$ is $U(1)$, the supersymmetric generalization of the first term in \eqref{3danomaly} is
\begin{align}
\int d^4\theta (\Sigma + \bar{\Sigma}) C_{IJK}(Y,\bar{Y}) Y^I (D_{\alpha} Y^J) (\bar{D}^{\alpha} \bar{Y}^K) \ , \label{superano1}
\end{align}
and the second term is
\begin{align}
\int d^4\theta (\Sigma + \bar{\Sigma}) C_{I}(Y,\bar{Y}) Y^I D_{\alpha} \bar{D}^{\alpha} V \ , \label{superano2}
\end{align}
where $V$ is a real vector superfield. Although we have not included it for simplicity, the R-anomaly proportional to $i(\Sigma-\bar{\Sigma})$ is also possible.

We have discussed that local counterterms introduce an additional contribution to the Weyl anomaly. When the local counterterms are chosen arbitrarily, we argued that they give Class 3 ambiguities. In particular, replacing $C_{IJK} (Y,\bar{Y})$ and $C_{I}(Y,\bar{Y})$ with $c_{IJK}(Y,\bar{Y})$ and $c_I(Y,\bar{Y})$ in \eqref{superano1} \eqref{superano2} and computing the Weyl variation, we obtain the $\mathcal{N}=2$ supersymmetric version of Class 3 ambiguities discussed in section \ref{sec2.2}.

One more interesting contribution to the Weyl anomaly comes from the supersymmetric Chern-Simons counterterms discussed in \cite{Closset:2012vg}\cite{Closset:2012vp}. Within R-symmetric $\mathcal{N}=2$ supergravity, they showed three-possible supersymmetric Chern-Simons counterterms. Among them, the gravitational Chern-Simons term is Weyl invariant by itself, so it does not lead to any Weyl anomaly, while Z-Z Chern-Simons term and Flavor-R Chern-Simons term do show the Weyl anomaly.

The bosonic part of the Z-Z Chern-Simons counterterm in component is
\begin{align}
{W}_{ZZ} =  -\frac{k_{ZZ}}{4\pi} \int d^3x \sqrt{|\gamma|} \left(\epsilon^{\mu\nu\rho}(a^R_\mu-\frac{1}{2}v_\mu) \partial_\nu(a^R_\rho-\frac{1}{2}v_\rho) + \frac{1}{2} H R + \cdots \right)
\end{align}
Here, $a^R_\mu$ is the vector source for the R-current, and $v_{\mu}$ is the vector source for the central charge current. When they are conserved, they do not give any Weyl anomaly  as a part of Class 3 ambiguities. On the other hand, $H$ is the source for dimension $2$ scalar operator (called $J^{(Z)}$ in \cite{Closset:2012vg}\cite{Closset:2012vp}) in the central charge current multiplet, so this is nothing but $\mathcal{I}$ term in \eqref{class3massive}. The counterterm is not Weyl invariant, and it induces the extra contribution to the Weyl anomaly as in \eqref{3amb}, which may or may not be cancelled from the other terms such as $k$ term in the Weyl anomaly that had existed before adding the Chern-Simons counterterm.

The bosonic part of the flavor-R Chern-Simons counterterm in component is 
\begin{align}
{W}_{fr} =  -\frac{k_{fr}}{2\pi} \int d^3x \sqrt{|\gamma|} \left(\epsilon^{\mu\nu\rho}a^f_\mu \partial_\nu(a^R_\rho-\frac{1}{2}v_\rho) + \frac{1}{4} \sigma R -DH \cdots \right) \ ,
\end{align}
where $a^f_\mu$ is the vector source for the flavor symmetry current, and $D$ and $\sigma$ are scalar sources for dimension $1$ and $2$ operators in the current supermultiplet \cite{Closset:2012vg}\cite{Closset:2012vp}.
When the flavor symmetry is conserved, the first Chern-Simons term do not contribute to the Weyl anomaly, but when it is not conserved, then non-zero vector beta functions will give terms similar to \eqref{cssmb} in the Weyl anomaly. 
Furthermore, the $R\sigma$ term and $DH$ term are $\mathcal{B}_{Mm}$ and $\mathcal{I}$ term in the Weyl anomaly, which give the extra contribution as in \eqref{3amb}.
 These terms may or may not be cancelled from the original Weyl anomaly terms such as $k$ term before adding the Chern-Simons counterterm.

\subsection{Holography} \label{sec4.3}
Our final example is the holographic computation of Schwinger vacuum functional. From the AdS/CFT correspondence, we identify Gubser-Klebanov-Polyakov-Witten free energy of the gravitational system in $d+1$ dimension as the Schwinger vacuum functional of the dual $d$-dimensional quantum field theory \cite{Gubser:1998bc}\cite{Witten:1998qj}. In the definition of the Gubser-Klebanov-Polyakov-Witten free energy, the space-dependent sources in field theory direction are naturally encoded as the boundary conditions for the bulk fields at the AdS boundary. 

In the AdS/CFT correspondence, the extra radial direction is understood as the renormalization group scale. The renormalization of the Schwinger functional is realized by the holographic renormalization of the Gubser-Klebanov-Polyakov-Witten free energy. The holographic renormalization group has been successful in deriving the holographic Weyl anomaly \cite{Henningson:1998gx}, holographic $c$-theorem \cite{Akhmedov:1998vf}\cite{Alvarez:1998wr}\cite{Girardello:1998pd}\cite{Freedman:1999gp}\cite{Sahakian:1999bd}\cite{Myers:2010tj} as well as the holographic equivalence between scale invariance and conformal invariance \cite{Nakayama:2009fe}\cite{Nakayama:2010wx}\cite{Nakayama:2010zz}. 

In this section, we would like to understand how our general framework of local renormalization group analysis and Weyl anomaly in $d=1+2$ dimensional quantum field theories fit with the holographic computation in $d=1+3$ dimensional effective semiclassical gravity.
We do not assume a particular string realization of the AdS/CFT correspondence, but we may apply the following argument to known holographic examples in string theory.

In order to obtain our new trace anomaly terms, we need to break the parity.
The simplest parity violating terms in the $d=1+3$ dimensional bulk can be obtained by topological $\theta$ terms for bulk gauge fields as well as by the gravitational $\theta$-term (Pontryagin-Hirzebruch term):
\begin{align}
S_f = \int d^4x \sqrt{|g|} \epsilon^{ABCD} \theta_f \mathrm{Tr} (F_{AB} F_{CD}) 
\end{align}
\begin{align}
S_g = \int d^4x \sqrt{|g|} \epsilon^{ABCD} \theta_g R_{AB}^{\ \ EF} R_{CDEF} \ . 
\end{align}
In this subsection earlier Latin indices $AB\cdots$ denote $d=1+3$ dimensional tensor indices.
These terms are equivalent to boundary Chern-Simons interaction after integration by part in the radial direction, and they give local contributions to the Gubser-Klebanov-Polyakov-Witten partition function \cite{Witten:2003ya}.

Thus we can easily obtain the contribution to the Weyl anomaly from these parity violating terms in the bulk action. First of all, the gravitational $\theta$-term  (Pontryagin-Hirzebruch term) does not produce any Weyl anomaly. The only effect is that we introduce a parity violating conformal invariant contact term in the two-point function of the energy-momentum tensor \cite{Witten:2003ya}\cite{Maldacena:2011nz}. On the other hand, the bulk gauge $\theta$-term does introduce the Weyl anomaly essentially by the same mechanism discussed in section \ref{sec2.2}. When there exist the vector beta functions $\hat{\rho}_I D^\mu g^I$ for the operator dual to $A_{\mu}$ appearing in the Chern-Simons interaction, then the contribution to the Weyl anomaly is
\begin{align}
A_{\mathrm{anomaly}} = \theta_f\epsilon^{\mu\nu\rho} \hat{\rho}_I f_{\mu\nu} D_{\rho} g^I \ .
\end{align}
Since $\theta_f$ is physical up to $2\pi$ integer shift, the effect cannot be removed by Class 3 ambiguity with a local counterterm, which must be integer shift of Chern-Simons term, and this essentially gives an existing proof of our Weyl anomaly terms in holography whenever $\theta_{f}$ is non-zero up to $2\pi$ integer shift.

In the bulk gravity, the vector beta functions near the conformal fixed point are understood as follows.  We use the Poincar\'e coordinate near the AdS boundary with metric $ds^2 = g_{AB} dx^A dx^B =  \frac{dz^2 + dx_\mu dx^\mu}{z^2}$. The non-trivial vector beta function means that the vector field $A_{\mu}$ is related to the scalar fields $\Phi^I$ dual to the boundary operator $O_I$ as
\begin{align}
A_\mu(z,x_\mu) = (\log z) \rho_I D_\mu \Phi^I(z,x_\mu) \ ,
\end{align}
where $\Phi^I(z,x_\mu)$ is slowly varying in the radial $z$ direction.
It is not so obvious that such a relation is compatible at the exact conformal fixed point with the AdS isometry. This is related to the question if we can have non-zero vector beta functions at the conformal fixed point, and it is not particular to AdS/CFT correspondence. One should notice, however, the bulk vector fields $A_{B}$ must be Higgesed \cite{Nakayama:2012sn}\cite{Nakayama:2013is} in order to obtain non-zero vector beta functions, breaking the conservation of the dual operator $J_\mu$. As discussed in section \ref{sec4.1}, generically the vector beta function is non-zero slightly away from the fixed point, and therefore the induced Weyl anomaly does not vanish.

We can also consider the parity violating terms which do not immediately give the local contribution to the Gubser-Klebanov-Polyakov-Witten functional. For instance, the bulk axion interaction
\begin{align}
S = \int d^4x \sqrt{|g|} \epsilon^{ABCD} \Theta_f(\Phi^I) \mathrm{Tr} (F_{AB} F_{CD}) 
\end{align}
will give non-zero contribution to the parity violating Weyl anomaly at the higher order in holographic computations (possibly with bulk loop factors).

It is possible to give holographic interpretations to various ambiguities discussed in previous sections. Class 1 ambiguity is given by the gauge transformation in the bulk. For simplicity, let us consider the $U(1)$ gauge field $A=A_B dx^B$ in the bulk. Let us also assume we have a charged scalar field $\Phi$ in the bulk so the gauge symmetry acts as
\begin{align}
\Phi &\to e^{i\Lambda}\Phi \cr
A& \to A + d\Lambda \ .
\end{align}
As discussed in \cite{Nakayama:2012sn}\cite{Nakayama:2013is}, this gauge transformation gives the holographic realization of Class 1 ambiguity when $\Phi$ has a non-trivial vacuum expectation value. For example, the bulk field configuration
\begin{align}
\Phi &= \gamma z^{i\alpha} \cr
A & = 0 
\end{align}
which is interpreted as $\beta^g = i\alpha \gamma g$ and $v = 0$ in the dual field theory is gauge equivalent to
\begin{align}
\Phi &= \gamma \cr
A & = \frac{\alpha dz}{z} \ ,
\end{align}
which is interpreted as $\beta^g = 0$ but $v = \alpha$ in the dual field theory. In both cases, the covariant derivative $zD_z \Phi = i\alpha \Phi$ in the radial direction is interpreted as the gauge invariant $B^I$ function of the dual field theory.

Class 2 ambiguity in holography is the scheme change of the bulk-boundary correspondence. The simplest example is the target space diffeomorphism for bulk scalar fields $\Phi^I \to \tilde{\Phi}^I (\Phi)$. This is nothing but the scheme change of the scalar coupling constants \eqref{scalarred} discussed in section \ref{sec2.2}. Other more involved field redefinitions in the bulk are possible such as $A_A \to A_A + r_I D_A \Phi^I$, which must be comparable with \eqref{vectorred}. In some cases, we may use these field redefinitions to make the gravitational action canonical such as the one in the Einstein frame, where energy-condition can be naturally applied, but the availability of such a choice may not be guaranteed in more complicated situations. Such ambiguities, in particular in relation to unitarity, are important issues begging for further studies in holography (see e.g. \cite{Myers:2010tj}).

Finally the holographic realization of Class 3 ambiguity is given by adding boundary counterterms, which is also understood as the bulk total derivative terms. We have already mentioned the effect of the boundary Chern-Simons terms above. 
When the coefficient is suitably quantized, they can be removed by the counterterms.
Another example would be the parity breaking interaction term
\begin{align}
\int d^4x \sqrt{|\gamma|} \epsilon^{ABCD} c_{IJKL} D_{A} \Phi^I D_{B} \Phi^J D_C\Phi^K D_D \Phi^L \ .
\end{align} 
When the scalar coupling constant has a non-zero beta functions
\begin{align}
zD_z \Phi^I \sim B^I
\end{align}
near the boundary, it is easy to see that $z$ integration gives rise to the logarithmic divergence near the boundary and we have the induced holographic Weyl anomaly
\begin{align}
\delta A_{\mathrm{anomaly}} = B^{L}c_{IJKL} \epsilon^{\mu\nu\rho} D_{\mu} g^I D_{\nu} g^J D_{\rho} g^K \ ,
\end{align}
which is comparable with the field theory Class 3 ambiguity \eqref{class3trace}. 

\section{Discussions} \label{sec5}
In this paper, we have discussed the consistency conditions and ambiguities in local renormalization group in most generic quantum field theories in $1+2$ dimension within power-counting renormalization scheme. We have argued that the consistency conditions from the local renormalization group require various non-trivial transversality conditions on beta functions and various tensors that appear in the trace anomaly. We have performed modest checks of these conditions in examples including supersymmetric field theories and holography.

As is the case with the other anomalies in different dimensions, the anomaly we have discussed in this paper must remain the same under the duality transformation up to ambiguities we have thoroughly discussed. In addition, it must satisfy the matching condition under the renormalization group flow. Therefore we may be able to use our new Weyl anomaly in $1+2$ dimensions for novel checks of the dualities proposed in the literature. For instance, $S$ in \eqref{massanomaou} is nothing but the operator product expansion coefficients of $O^{(m)}$ at the conformal fixed point and they must agree between duality pairs.

With respect to the anomaly matching, it would be interesting to construct the Wess-Zumino action as the integrated form of the anomaly in contrast to the infinitesimal variation we have discussed in this paper.  After all, the Wess-Zumino conditions guarantee that the integration is possible.
The integrated Weyl anomaly in even dimensions are studied as dilaton effective action in \cite{Schwimmer:2010za}\cite{Komargodski:2011vj} at the conformal fixed point in relation to proving the $a$-theorem in $1+3$ dimension. The complete dilaton effective action off criticality incorporating the space-time dependent coupling constant contribution was obtained in \cite{JO}\cite{R} (see \cite{Fortin:2012hn}\cite{Nakayama:2013is} for related computations). It is possible to apply the same technique here in $1+2$ dimension. We only note, however, that the parity violating contribution to the on-shell dilaton scattering is trivial due to the Bose symmetry (see \cite{Nakayama:2013is} for a related remark in $d=1+3$ dimension).

In this paper, we have not addressed the question if the conjectured F-theorem \cite{Myers:2010tj}\cite{Jafferis:2011zi} could be understood from the consistency conditions of the renormalization group (and probably with other assumptions such as unitarity). While our consistency conditions give various constraints on the renormalization group flow, we have not so far obtained the equation analogous to \eqref{atheorem} valid in even space-time dimensions. Probably, we should study the properties of the partition function itself by integrating the Weyl transformation explicitly.

\section*{Acknowledgements}
The author would like to thank H.~Osborn for discussions and sharing his note.
He would like to thank CERN theory division and APCTP for hospitality  where the current research was developed. He in particular thanks organizers of the wonderful workshops there.
This work is supported by Sherman Fairchild Senior Research Fellowship at California Institute of Technology  and DOE grant DE-FG02-92ER40701.

\appendix
\section{Inclusion of cosmological constant} \label{appa}

The introduction of the cosmological constant in local renormalization group analysis is possible but does not lead to any new interesting results. Let us see this in $d=1+2$ dimension without any other mass parameters nor current operators. This is just for simplicity, and the similar results apply in most generalities with various dimensionful couplings even in other space-time dimensions.
  
The local renormalization group operator with the cosmological constant is given by
\begin{align}
\Delta_{\sigma} = \int d^3x \sqrt{|\gamma|} &\left(2\sigma \gamma_{\mu\nu} \frac{\delta}{\delta \gamma_{\mu\nu}} + \sigma \beta^I \frac{\delta}{\delta g^I} + \sigma(3-\gamma_{\Lambda} ) \Lambda \frac{\delta}{\delta \Lambda} \right. \cr
 &- \left. \sigma \epsilon^{\mu\nu\rho} \hat{c}_{IJK} \partial_\mu g^I \partial_\nu g^J \partial_\rho g^K \frac{\delta}{\delta \Lambda} + \epsilon^{\mu\nu\rho}  \hat{k}_{IJ} \partial_\mu \sigma \partial_\nu g^I \partial_\rho g^J  \frac{\delta}{\delta \Lambda} \right) \ , \label{ccvar}
\end{align}
with totally antisymmetric $\hat{c}_{IJK}$ and $\hat{k}_{IJ}$,
and we assume that the renormalized Schwinger vacuum functional is annihilated by $\Delta_{\sigma}$. 
We do not need to address the ``anomalous variation" of the vacuum functional with respect to the local renormalization group because the variation for the cosmological constant on the local functional gives the same effect. Indeed the trace identity from \eqref{ccvar} is
\begin{align}
T^{\mu}_{\ \mu} = \beta^I O_I - (\hat{c}_{IJK} + \partial_I \hat{k}_{JK} ) \epsilon^{\mu\nu\rho}  \partial_\mu g^I \partial_\nu g^J \partial_\rho g^K +  (3-\gamma_{\Lambda}) \Lambda \ , \label{tracecc}
\end{align}
which should be compared with \eqref{traceiden} and \eqref{traceanomalymassless}. In particular, the second term is what was called $A_{\mathrm{anomaly}}$, but here it is obtained without the explicit anomalous variation.

Let us consider the additional Class 1 consistency conditions  due to the cosmological constant from $[\Delta_{\sigma}, \Delta_{\tilde{\sigma}}] = 0 $.
They are given by 
\begin{align}
3\hat{c}_{IJK} \beta^K + \mathcal{L}_\beta \hat{k}_{IJ}  +\gamma_\Delta \hat{k}_{IJ} &= 0 \ \cr
\hat{k}_{IJ} \beta^J &= 0 \label{cccon}
\end{align}
These are equivalent to the equation \eqref{consistenccc} by identifying 
\begin{align}
C_{IJK} = \hat{c}_{IJK} + \partial_{[I} \hat{k}_{JK]} \ , \label{invc}
\end{align}
which is motivated by the trace identity \eqref{tracecc}, 
if we assume that the anomalous dimension of the cosmological constant is zero: $\gamma_{\Lambda} = 0$. This vanishing of the anomalous dimension is a reasonable assumption in our situation because the dimension of the identity operator, which must be zero conventionally, after all determines what we mean by scaling transformation.\footnote{In contrast, if we allow other dimension zero operators as in non-compact conformal field theories, the situation can become more subtle. See \cite{Osborn:1991gm} for two-dimensional discussions for such a case.} Due to the absence of the anomalous variation, there is no Class 2 consistency condition. Instead it was encoded in Class 1 as shown above.

At first sight, we have more freedom than the discussion in section \ref{sec2.2} because of the additional term $\hat{k}_{IJ}$. The necessity of such additional terms can be seen, for example, from Class 2 ambiguity with $\mathcal{D} = \int d^3x \sqrt{|\gamma|}  c_{IJK} \epsilon^{\mu\nu\rho} \partial_\mu g^I \partial_\nu g^J \partial_\rho g^K\frac{\delta}{\delta \Lambda}$. From the variation $\delta \Delta_{\sigma} = [\Delta_{\sigma}, \mathcal{D}]$, we have to allow the scheme dependence 
\begin{align}
\delta \hat{c}_{IJK} &= \mathcal{L}_\beta c_{IJK}  \cr
\delta \hat{k}_{IJ} &= c_{IJK} \beta^K \ ,
\end{align}
which satisfies the consistency conditions \eqref{cccon}.
However, under the same scheme change, the invariant combination $C_{IJK}$ defined in \eqref{invc} transforms as
\begin{align}
\delta C_{IJK}  = \beta^L \partial_{[L}C_{IJK]} \ ,
\end{align}
which is equivalent to the first line in \eqref{class3trace} obtained as Class 3 ambiguity there. In this way, we do not obtain any new physically interesting constraint or ambiguity in the local renormalization group analysis with the addition of the cosmological constant.

\section{Convention} \label{appb}
We mostly follow the convention and notation used in \cite{JO}, which is slightly different from the ones used in \cite{Nakayama:2013is}. We here list minor difference and the convention implicit in \cite{JO}. 

Our definition of the functional differentiation is defined with respect to the  explicit volume measure:
\begin{align}
\frac{\delta f(y)}{\delta f(x)} &= \frac{1}{\sqrt{|\gamma|}} \delta^{(d)}(x-y)
\end{align}
or
\begin{align}
\frac{\delta}{\delta f(x)} \int d^dx \sqrt{|\gamma|} f(x) g(x) &= g(x) \ .
\end{align}
The definition differs from \cite{JO} by the factor $\sqrt{|\gamma|}$, but it does not affect most of our formulae.

Our anti-symmetrization symbol $[IJK \cdots ]$ and symmetrization symbol $(IJK\cdots)$ in tensor indices contain the combinatoric factor so that $A_{[IJL\cdots]}$ and $S_{(IJK\cdots)}$ represent the antisymmetric or symmetric component of the corresponding tensor $A_{IJK\cdots}$ or $S_{IJK\cdots}$. For instance
\begin{align}
A_{[IJK]} = \frac{1}{6}\left(A_{IJK} - A_{IKJ} - A_{JIK} + A_{JKI} +A_{KIJ} -A_{KJI}\right) \cr
S_{(IJK)} = \frac{1}{6}\left(S_{IJK} + S_{IKJ} + S_{JIK} + S_{JKI} + S_{KIJ} + S_{KJI} \right) \ .
\end{align}

Our convention of the Levi-Civita tensor is as follows.
We first define the totally anti-symmetric Levi-Civita symbol $\epsilon_{abc\cdots}$ in the Euclidean signature as the $c$-number
\begin{align}
\epsilon_{123 \cdots d} = 1
\end{align}
and $\pm 1$ depending on odd or even under permutations.
With the vielbein $\gamma_{\mu\nu} = e^{a}_{\mu}e^{b}_{\nu} \delta_{ab}$ for the Riemannian metric, 
we define the Levi-Civita tensor as
\begin{align}
\epsilon_{\mu\nu\rho \cdots} = i \epsilon_{abc \cdots} e^{a}_{\mu} e^{b}_{\nu} e^c_{\rho} \cdots \ .
\end{align}
Note that the imaginary unit $i = \sqrt{-1}$ here can be attributed to the Wick rotation
so that the Levi-Civita tensor in the Lorentzian signature is real, which guarantees  the reality of the effective action in the Lorentzian space-time.

Our curvature convention is the same as the one used in \cite{Nakayama:2013is}, or $s_1 = s_2 = s_3 = +$ in the Misner-Thorne-Wheeler convention.
Under the infinitesimal Weyl transformation 
\begin{align}
\delta_{\sigma} \gamma_{\mu\nu} = 2\sigma \gamma_{\mu\nu} 
\end{align}
we have
\begin{align}
\delta_{\sigma} R &= -2\sigma R - 2(d-1) D^2 \sigma \cr
\delta_{\sigma} D^2 &= -2\sigma D^2 + (d-2) (\partial_\mu \sigma) D^\mu  \ .
\end{align}
Here we assume $D^2 = D^\mu \partial_\mu$ acts on scalar fields.

\end{document}